\documentclass{IEEEtran}
\usepackage{cite}
\usepackage{amsmath,amssymb,amsfonts}
\usepackage{graphicx}
\usepackage{textcomp,nicefrac}
\usepackage[switch, columnwise]{lineno}

\def\BibTeX{{\rm B\kern-.05em{\sc i\kern-.025em b}\kern-.08em
T\kern-.1667em\lower.7ex\hbox{E}\kern-.125emX}}
\begin{document}
%\linenumbers
\bstctlcite{IEEEcontrol}
\makeatletter
\def\bstctlcite{\@ifnextchar[{\@bstctlcite}{\@bstctlcite[@auxout]}}
\def\@bstctlcite[#1]#2{\@bsphack
  \@for\@citeb:=#2\do{%
    \edef\@citeb{\expandafter\@firstofone\@citeb}%
    \if@filesw
      \immediate\write\csname #1\endcsname{\string\citation{\@citeb}}%
    \fi}%
  \@esphack}
\makeatother

\title{ CHARMS250V1: A Cryogenic Front-End ASIC for Low-Noise Readout of Charge or Light Signals in Noble Liquid Detectors}
% \author{First A. Author, \IEEEmembership{Fellow, IEEE}, Second B. Author, and Third C. Author, Jr., \IEEEmembership{Member, IEEE}
% \thanks{This paragraph of the first footnote will contain the
% date on which you submitted your paper for review. It will also contain
% support information, including sponsor and financial support acknowledgment.
% For example, ``This work was supported in part by the U.S. Department of
% Commerce under Grant BS123456.'' The next few paragraphs should contain the
% authors' current affiliations, including current address and e-mail. For
% example, F. A. Author is with the National Institute of Standards and
% Technology, Boulder, CO 80305 USA (e-mail: author@ boulder.nist.gov).}
% \thanks{S. B. Author, Jr., was with Rice University, Houston, TX 77005 USA. He is
% now with the Department of Physics, Colorado State University, Fort Collins,
% CO 80523 USA (e-mail: author@lamar.colostate.edu).}
% \thanks{T. C. Author is with
% the Electrical Engineering Department, University of Colorado, Boulder, CO
% 80309 USA, on leave from the National Research Institute for Metals,
% Tsukuba, Japan (e-mail: author@nrim.go.jp).}}
\author{Prashansa~Mukim, Shanshan~Gao, Hucheng~Chen, Grzegorz~W.~Deptuch, Jack~Fried, Niccol\`{o}~Gallice,  Dominik~Gorni, Lingyun~Ke, Giovanni~Pinaroli, Sergio~Rescia, and Chao Zhang
\thanks{Prashansa~Mukim, Grzegorz~W.~Deptuch, Jack~Fried, Niccol\`{o}~Gallice, Dominik~Gorni, Giovanni~Pinaroli, and Sergio~Rescia are with the Instrumentation Division, Brookhaven National Laboratory, Upton NY 11973, USA. Email: pmukim@bnl.gov.}
\thanks{Shanshan~Gao, Hucheng~Chen, Lingyun~Ke, and Chao~Zhang are with the Physics Division, Brookhaven National Laboratory, Upton NY 11973, USA.}
\thanks{This manuscript has been authored by employees of Brookhaven Science Associates, LLC under Contract No. DE-SC0012704 with the U.S. Department of Energy.}%
}

\maketitle

\begin{abstract}
This work presents the design and measurement results of CHARMS250V1, a next-generation cryogenic front-end application specific integrated circuit (ASIC) developed using a 65\,nm CMOS process. It is designed for low-noise readout of charge or light signals generated in noble liquid detectors. CHARMS250V1 is a single channel prototype whose design in the future will be integrated within a 16-channel ASIC. It comprises pre-amplification and pulse shaping stages that provide programmable charge gain between 4.3\,mV/fC and 25\,mV/fC, peaking time programmable between 250\,ns and 2\,\textmu{}s, and programmable baseline of either 200\,mV or 900\,mV for processing of unipolar and bipolar signals, respectively. 
%It is designed for operation at temperatures ranging from room temperature (RT) down to liquid nitrogen temperature (LNT), i.e., 77\,K, with large capacitance detectors (up to hundreds of picofarads). 
% The charge gain for each channel can be set to 60, 100, 180 or 320 for readout of input signals up to 300\,fC, 180\,fC, 100\,fC and 56\,fC, respectively, while the pulse peaking time can be set to 0.25\,\textmu{}s, 0.5\,\textmu{}s, 1.0\,\textmu{}s or 2.0\,\textmu{}s. 
% Simulations with both room and cryogenic temperature transistor model parameters indicate that CHARMS250 provides a highly linear voltage readout (INL $<$ 0.05\%) for input charge up to 300\,fC with a power consumption of 6-11\,mW per channel. A prototype ASIC comprising of a single-channel has been submitted for fabrication.
%and baseline equivalent noise charge (ENC) less than 500 electrons at LNT for all peaking times and detector capacitcance of 160\,pF.
 Measurements indicate that CHARMS250V1 demonstrates stable operation across all programmable settings from 77\,K to 300\,K, per channel power consumption of 6-12\,mW, less than 1\% integral non-linearity and equivalent noise charge of 980\,$e^-$ at 77\,K for a peaking time of 250\,ns and 200\,pF detector capacitance. 
\end{abstract}


\section{Introduction}
\label{sec:introduction}
% Cold CMOS electronics has enabled low-noise measurements in liquid argon TPCs (protoDUNE, microBooNE, DUNE). LArASIC, which is a pre-amplification and shaping filter analog front-end ASIC, will be used in DUNE Phase-I far detector modules. However it is in the 180 nm node. Next generation liquid argon TPC experiments need ASICs in advanced technology nodes. This work presents the design and measurement results of the first prototype cryogenic ASIC in 65 nm designed at BNL, i.e., CHARMS250V1, that contains a single analog front-end channel. Future work is aimed at integrating 16-channels into a single ASIC.

Cryogenic complementary metal-oxide-semiconductor (CMOS) front-end electronics have enabled low-noise measurements in liquid argon time projection chambers (LArTPCs)~\cite{Radeka_2011}\cite{CHEN2023167571} and have been successfully deployed in experiments such as the Micro Booster Neutrino Experiment (MicroBooNE)~\cite{Acciarri_2017}\cite{microboone2017}, the Short-Baseline Near Detector (SBND)~\cite{gao2019lownoisecoldelectronics}, and the Prototype Deep Underground Neutrino Experiment (ProtoDUNE)~\cite{protoDUNE2017}\cite{protoDUNE2022}\cite{Abbaslu_2026}. The Liquid Argon Application-Specific Integrated Circuit (LArASIC)~\cite{larasic_2010}\cite{TCAS2025}, which provides 16 channels of low-noise charge amplification and pulse shaping with programmable gain and peaking time, has been adopted as the analog front-end ASIC for charge readout in the Deep Underground Neutrino Experiment (DUNE)~\cite{Abi_2020}\cite{Abud_2024} Phase-I far detector modules. LArASIC is implemented in a 180\,nm CMOS process and has demonstrated excellent linearity, low equivalent noise charge (ENC), and reliable operation in liquid argon, i.e. at 89\,K.

Next-generation noble liquid detectors, including the DUNE Phase-II far detector modules and the liquid argon electromagnetic calorimeter proposed for the A Lepton-Lepton collider Experiment with Granular Read-Out (ALLEGRO) detector concept at the Future Circular Collider electron-positron phase (FCC-ee)~\cite{FCC2025Feasibility}\cite{FCC2025FeasibilityVol2}, require a new generation of cryogenic front-end ASICs implemented in advanced technology nodes.
Moving to advanced technology nodes allows larger bandwidth, shorter shaping times, greater integration, and eases extensive digital programmability. Independently of these benefits, such a transition is necessary to maintain access to fabrication as the processes used for existing designs are discontinued.
%While ensuring continued fabrication by foundries in the next few decades, designs in advanced nodes can enable higher bandwidth, shorter shaping times, improved integration, and enhanced digital programmability. 
%while maintaining the low-noise cryogenic performance required for noble liquid detectors. 

This paper presents the design and measurement results of CHARMS250V1, the first cryogenic analog front-end ASIC prototype designed at Brookhaven National Laboratory (BNL) in a 65\,nm CMOS process. CHARMS250V1 contains a single analog front-end channel comprising a charge amplifier and pulse-shaping filter and its design builds upon the 180\,nm LArASIC chip. CHARMS250V1 provides programmable peaking times ranging from 250\,ns to 2\,\textmu{}s, compared with 500\,ns to 3\,\textmu{}s in LArASIC. This represents a two-fold reduction in the minimum peaking time without increasing the per-channel power consumption. CHARMS250V1 also introduces enhanced digital programmability of analog functions to improve robustness against process variations. These features include programmable global bias voltages and currents, as well as a per-channel baseline adjustment mechanism designed to compensate for device mismatches and preserve the input charge range. The serial peripheral interface (SPI) used for slow control in LArASIC has been replaced with a standard I\textsuperscript{2}C interface in CHARMS250V1.  The long-term goals of this work are to develop a 16-channel cryogenic front-end ASIC in the same CMOS technology and to build upon the existing CHARMS250V1 architecture to support an approximately 30-fold increase in the full-scale input charge range, targeting the requirements of next-generation noble liquid calorimeters.

\section{Architecture of CHARMS250V1}
\label{sec:arch}

%\textbf{Single channel block diagram - describe it, , I2C interface, BGR, temperature sensor, 1.8 V and 1.2 V power supplies, Vdda, Vddo, CAs similar to LArASIC, but implement IRQI subtraction. Local generation of IRQI. Thick oxide input transistor. Shaping filter - scaled components for peaking time between 250 ns and 2 us. similar SE and SEDC buffers. Show IRQI subtraction circuitry.}

\begin{figure}
    \centering
    \includegraphics[width=\linewidth]{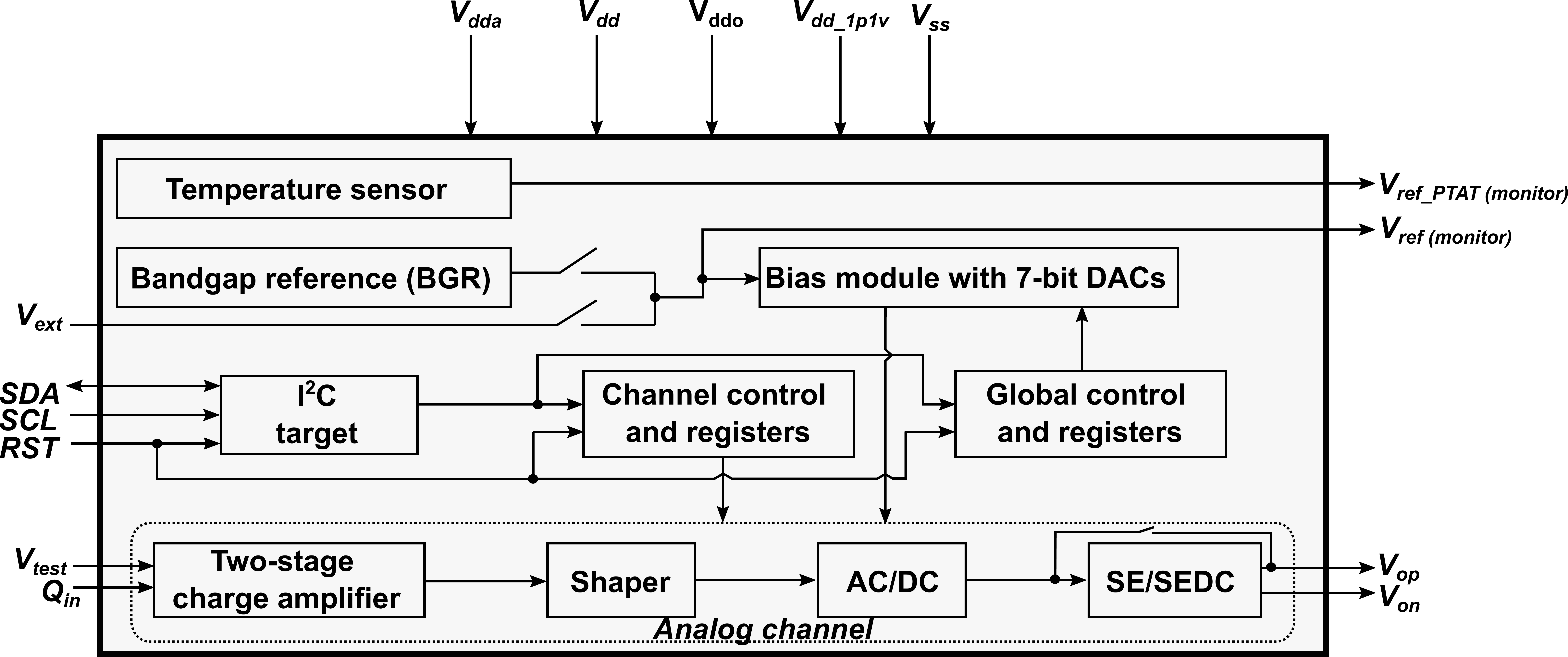}
    \caption{Block diagram of the CHARMS250V1 architecture.}
    \label{fig:charms-block}
\end{figure}

The general architecture of CHARMS250V1 is shown in Fig.~\ref{fig:charms-block}. It comprises a single analog front-end channel that includes two stages of charge amplifiers~\cite{CSAOConnor}. 
In response to an input charge \textit{Q\textsubscript{in}}, the first stage provides a fixed charge gain of 20, and the second stage provides a programmable charge gain of 3, 5, 9 or 16. The charge amplifiers are followed by a shaping filter~\cite{ShaperDesignGeronimo}
to produce semi-Gaussian voltage pulses with programmable output voltage to input charge gain of approximately 4.3\,mV/fC, 7.4\,mV/fC, 13\,mV/fC or 25\,mV/fC, with corresponding full-scale input charge range of 320\,fC, 200\,fC, 110\,fC or 60\,fC, respectively. The shaping filter converts the amplified current signals to voltage pulses with programmable peaking time $t_p$, of 0.25, 0.5, 1.0, or 2.0\,\textmu{}s, and provides anti-aliasing before digitization. 
The shaping filter output can be either AC- or DC-coupled to the output, and read out either directly, through a single-ended (SE) buffer on the pin \textit{V\textsubscript{op}}, or through a single-ended-to-differential converter (SEDC) buffer on the pins \textit{V\textsubscript{op}} and \textit{V\textsubscript{on}}. 
The SE and SEDC buffers are suitable for driving larger capacitive loads, and the SEDC buffer is introduced to reduce inter-channel crosstalk on transmitted signals. The output baseline voltage can be set to 200\,mV or 900\,mV for operation in charge collection and induction modes, respectively. The 200\,mV baseline allows processing of input charges of only negative polarity, whereas the 900\,mV baseline allows processing of bipolar input signals. The analog channel utilizes a 1.8\,V supply voltage between \textit{V\textsubscript{dd}} and \textit{V\textsubscript{ss}}, with the dedicated pins \textit{V\textsubscript{dda}} and \textit{V\textsubscript{ddo}} for  supplying separately the input transistor in the first charge amplification stage, and the output transistors in the SE and SEDC buffers, respectively, to avoid coupling and feeding of noise from other circuits into the input and output stages.

An integrated test capacitor, designed for a value of 200\,fF is available at the input of the channel to allow injection of external pulses through the pin \textit{V\textsubscript{test}} for testing and calibration of the channel.
All biasing circuits utilize a bandgap voltage reference (BGR), to ensure that the channels operate with near-constant current at temperatures ranging between room temperature (RT) and liquid nitrogen temperature (LNT), i.e., 77\,K. The BGR output can be monitored externally on the pin \textit{V\textsubscript{ref}}. For test purposes, the biasing circuits can alternatively be driven by an external reference voltage, \textit{V\textsubscript{ext}}. A temperature sensor is also included on the chip, and its output \textit{V\textsubscript{ref\_PTAT}} can be monitored externally.

CHARMS250V1 features enhanced digital programmability of all bias voltages and currents through 7-bit digital-to-analog converters (DACs), improving its robustness for operation at a wide range of temperatures ranging between RT and LNT. 
Slow control is provided through a standard I\textsuperscript{2}C interface, where the \textit{SCL} pin is used for transmitting the clock signal, the \textit{SDA} pin is used for bi-directional data transfer and there is the additional \textit{RST} pin, acting as a global reset for all registers in the I\textsuperscript{2}C block. Two blocks of the control logic are implemented to program the global registers for the bias module and the channel registers for the analog channel, respectively. The I\textsuperscript{2}C is synthesized with core transistors that utilize a 1.1\,V power supply \textit{V\textsubscript{dd\_1p1V}} and feature the minimum length of 90\,nm, which is 1.5$\times$ longer than the minimum length in this process. These precautions are used for long-term reliability, i.e. to increase robustness against hot carrier effect that is more pronounced at cryogenic temperatures~\cite{hce_Li2013}~\cite{hce_cryo}.

\section{Analog channel circuits of CHARMS250V1}

\begin{figure}
    \centering
    \includegraphics[width=\linewidth]{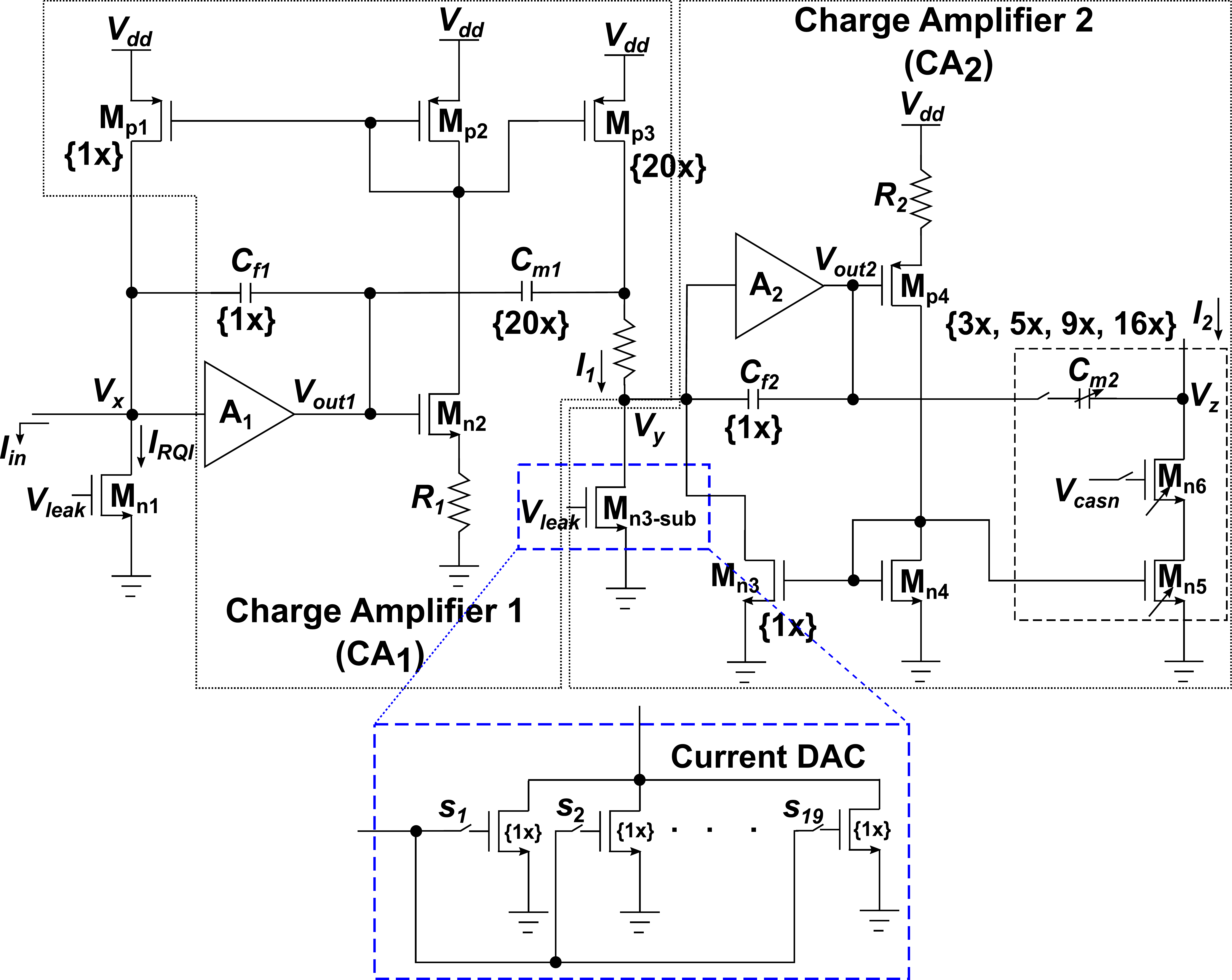}
    \caption{Circuit diagram of the charge amplification stages.
    %question about correctness, switches on RQI transistors, cannot leave gates floating while open, it is simplification shown, but it is not mentioned that this is a very sensitive deep subthreshold DAC so the gates needs to be grounded when the switches are off, also i have some doubts to how the gain switches are depicted at C2
    }
    \label{fig:charms-ca}
\end{figure}

Each channel of CHARMS250V1 contains two cascaded charge-amplification stages, CA\textsubscript{1} and CA\textsubscript{2}, as shown in Fig.~\ref{fig:charms-ca}. Both CA\textsubscript{1} and CA\textsubscript{2} utilize pole-zero cancellation~\cite{PoleZeroGrybos} and the  current-mirror-based continuous adaptive reset mechanism~\cite{GEMTPCASIC}~\cite{ResetOConnor}, which also stabilizes the DC operating point of the amplifiers. Due to the $20\times$ charge multiplication performed by CA\textsubscript{1}, the pre-bias reset quiescent current $I_{RQI}$ of CA\textsubscript{2} flowing through M\textsubscript{p3} is also multiplied by $20\times$, resulting in a shift of the DC-baseline at the output of the amplifier A\textsubscript{2}.
Device mismatches due to fabrication variations can introduce channel-to-channel dispersion of this baseline shift in a multi-channel implementation. Since a common baseline setting must accommodate the worst-case deviations across all channels, it leads to a reduction in the available signal swing and input charge range for other channels. To address this limitation, CHARMS250V1 incorporates a programmable baseline adjustment mechanism based on the $I_{RQI}$ subtraction, designed to be implemented independently in each channel of the future multi-channel ASIC, allowing the baseline adjustment per channel to compensate for channel-to-channel variations, particularly at cryogenic temperatures.
%and reducing the headroom for handling the signal swing. Thus, CHARMS250V1 targets a reduction mechanism for this drift that is robust to process mismatches and layout-induced non-idealities, especially at cryogenic temperatures.
%Add one more sentence about why it is greater at cryo
The pre-bias current of CA\textsubscript{2} is subtracted through the transistor M\textsubscript{n3-sub}, which is implemented as a current DAC with programmable $I_{RQI}$ subtraction between $0-19\times$ through switches $s_1$-$s_{19}$, as shown in Fig.~\ref{fig:charms-ca}. The current DAC transistors operate in deep subthreshold, with their gates connected to the ground terminal through a complementary set of switches, when given subtraction transistors are not used.

\begin{figure}
    \centering
    \includegraphics[width=\linewidth]{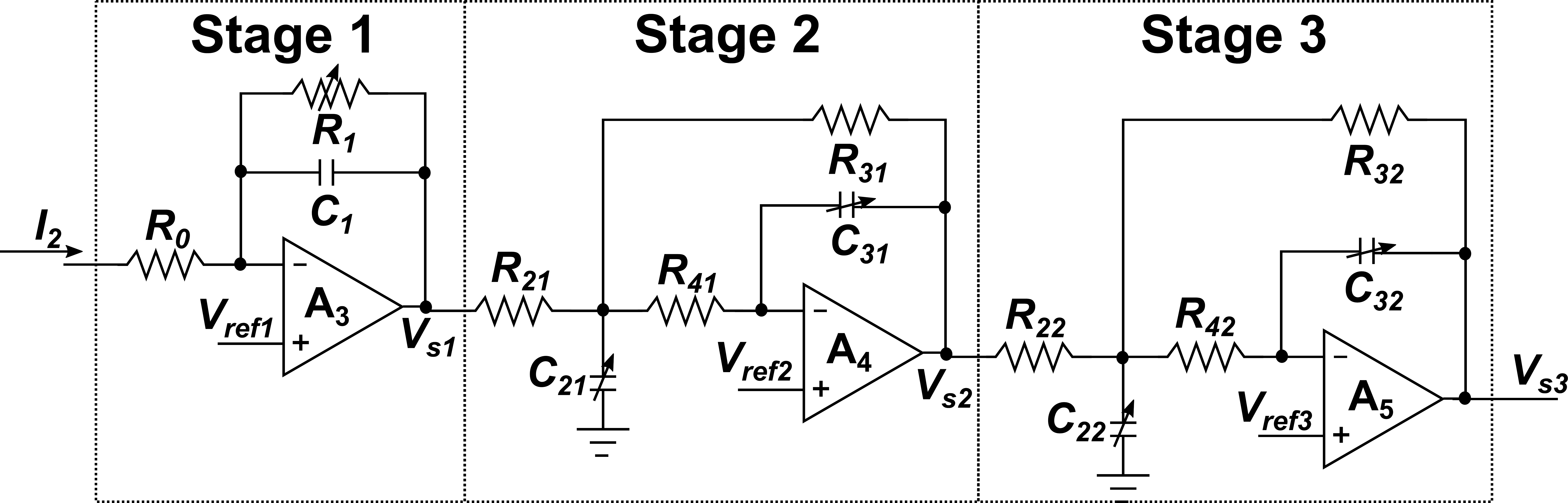}
    \caption{Circuit diagram of the shaping filter stages.}
    \label{fig:charms-shaping-filter}
\end{figure}

The input transistor of amplifier A\textsubscript{1} is a thick-oxide device, in order to minimize parallel noise originating from the high gate-leakage current in thin-oxide devices. A p-channel device is chosen for the input transistor due to its reduced 1/f noise contribution~\cite{larasic_2010}~\cite{2024Kiene}. It is sized to have a W{/}L=24\,mm\,/\,400\,nm, biased with 2\,mA current and has a transconductance of 48\,mS at RT. The gate capacitance of this device is 40\,pF, aiming at the minimization of the ENC for a detector capacitance of 100-200\,pF, within the given power budget. 
%Include one sentence about thick vs. thin oxide devices
The charge amplification stages are followed by a fifth-order semi-Gaussian pulse-shaping filter~\cite{OHKAWA197685}, shown in Fig.~\ref{fig:charms-shaping-filter}. %which converts amplified current pulses to symmetric voltage pulses. 
The first stage is a first-order low-pass filter (current to voltage converter), and is followed by a cascade of two second-order low-pass filter stages with the transfer function exhibiting one pair of complex conjugate poles. The resistors and capacitors marked as variable in Fig.~\ref{fig:charms-shaping-filter} provide programmability of the peaking time. Both the SE and SEDC buffer circuits employ a folded-cascode current-mirror-based operational transconductance amplifier with a rail-to-rail input stage and a push-pull output stage~\cite{TCAS2025}. 
%The SEDC buffer comprises of fully-differential input and output cores, and a differential output signal is generated by connection of its non-inverting output to the inverting input. It additionally requires a low-noise common-mode reference voltage, which is derived from the on-chip BGR and filtered locally within each channel. 
For the design of these circuits, simulations were performed with BSIM4 models~\cite{bsim4} with parameters calibrated at 89\,K through device measurements. 

\section{Measurement Results}
\label{sec:meas_results}
\begin{figure}
    \centering
    \includegraphics[width=\linewidth]{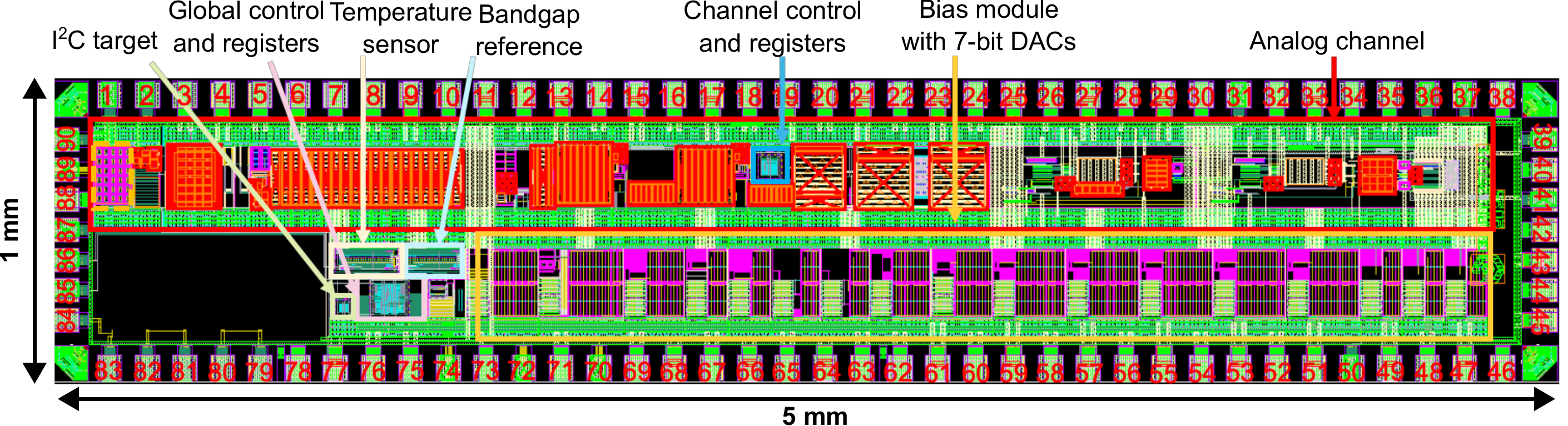}
    \caption{CHARMS250V1 mask layout.}
    \label{fig:chip-layout}
\end{figure}

\begin{figure}
    \centering
    \includegraphics[width=\linewidth]{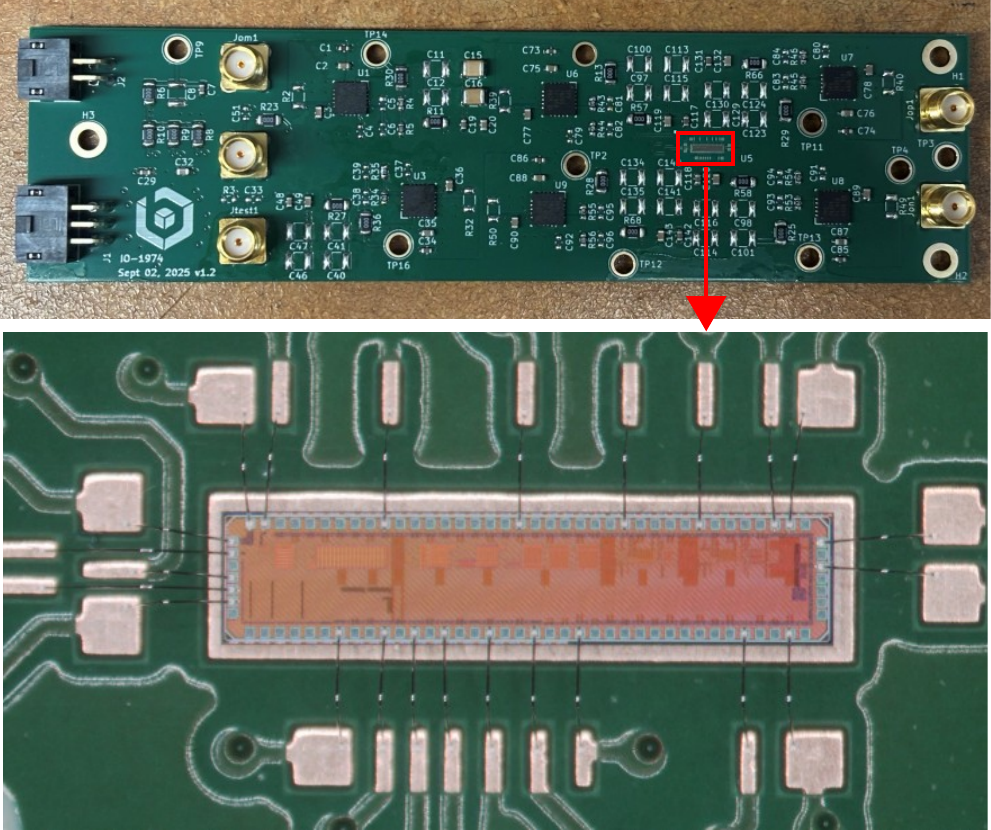}
    \caption{CHARMS250V1 test board and wire bonded die.}
    \label{fig:board-and-die}
\end{figure}

\begin{figure}
    \centering
    \includegraphics[width=\linewidth]{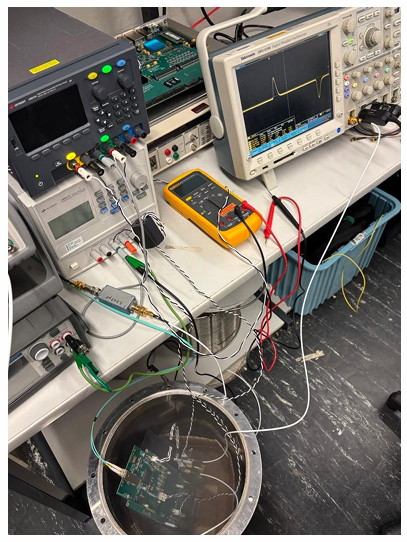}
    \caption{CHARMS250V1 LNT test setup.}
    \label{fig:LNT-test}
\end{figure}

An image of the mask layout of the fabricated CHARMS250V1 chip with dimension 5\,mm\,$\times$\,1\,mm is shown in Fig.~\ref{fig:chip-layout}. The test board with the chip wire-bonded to it is shown in Fig.~\ref{fig:board-and-die}. Measurements of the chip were performed at RT as well as LNT, with the LNT measurement test setup shown in Fig.~\ref{fig:LNT-test}. This figure shows an oscilloscope screen with the measured response of the chip when operated in the 900\,mV baseline setting. All measurements presented in this work were performed on a single randomly selected chip, without prior screening or performance-based selection, and are typical for the performance of CHARMS250V1. Chip-to-chip variations and the statistical spread of the measured parameters are not covered in this work.

%\textbf{Internal cap used for testing, calibrated against an external cap of value xx.}

% \begin{figure}
%     \centering
%     \includegraphics[width=\linewidth]{figures/CHARMS250_2p5x.pdf}
%     \caption{CHARMS250V1 die micrograph}
%     \label{fig:chip-picture}
% \end{figure}

\begin{figure}
    \centering
    \includegraphics[width=\linewidth]{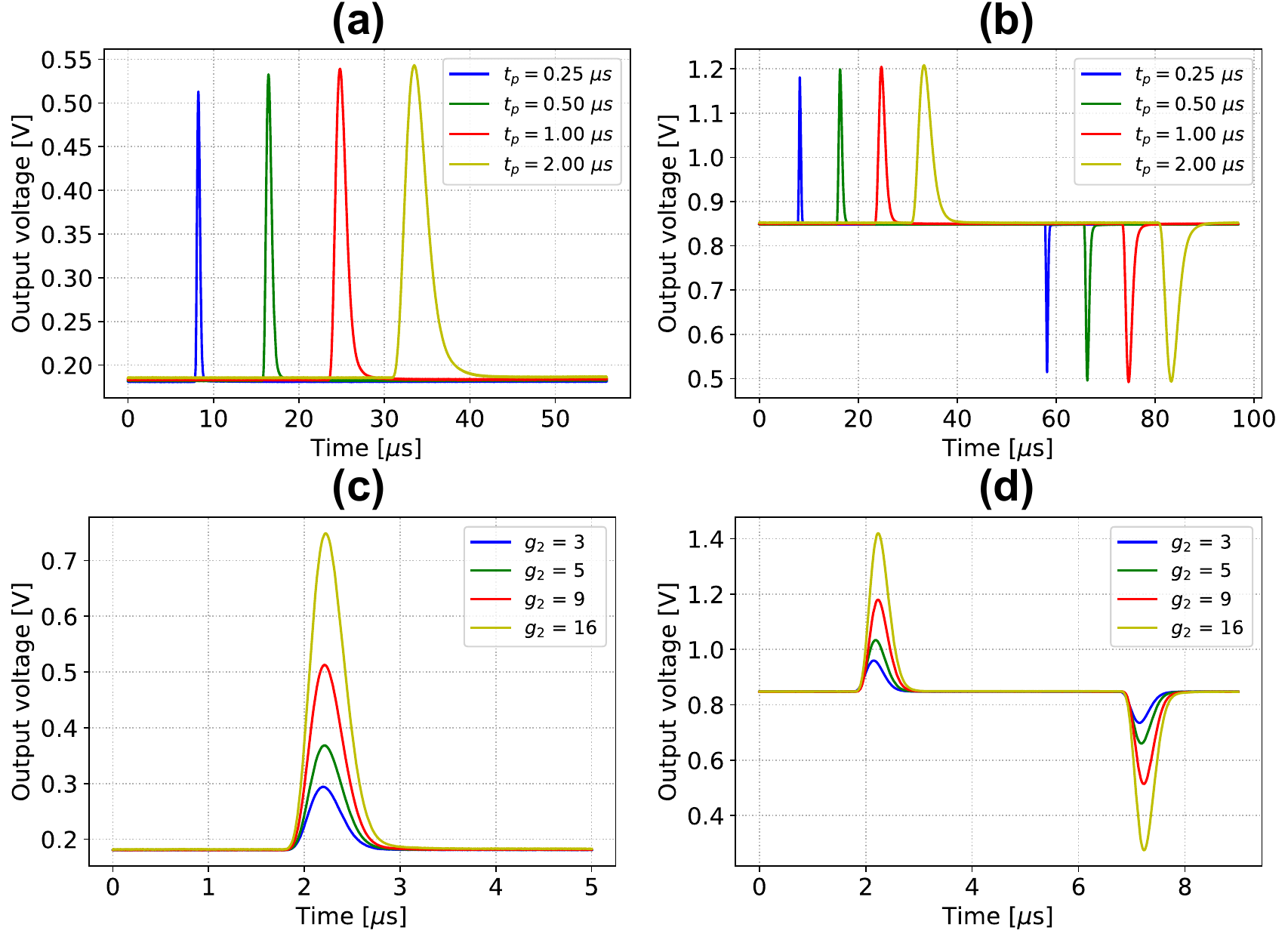}
    \caption{Measured impulse response at LNT for an input charge injection $Q_{in} = $25\,fC and (a) variable peaking time setting with 200\,mV baseline and $g_2$ = 9, (b) variable peaking time setting with 900\,mV baseline and $g_2$ = 9, (c) variable gain setting $g_2$, 200\,mV baseline and peaking time $t_p$ = 250\,ns, and (d) variable gain setting $g_2$, 900\,mV baseline and peaking time $t_p$ = 250\,ns.}
    \label{fig:LNT-waveforms}
\end{figure}

The measured impulse response of CHARMS250V1 at LNT for an injected charge $Q_{in}$ of 25\,fC for various settings is shown in Fig.~\ref{fig:LNT-waveforms}. Fig.~\ref{fig:LNT-waveforms}(a) and (b) show the responses for the four
programmable peaking-time settings, $t_p$, with the fixed second-stage charge amplifier gain ($g_2$) of 9, for the 200\,mV and 900\,mV baseline settings, respectively.
Figures~\ref{fig:LNT-waveforms}(c) and (d) show the measured impulse response at the four programmable second-stage gain ($g_2$) values, at the shortest peaking-time setting of 250\,ns, again for the baseline settings of 200\,mV and 900\,mV, respectively. An input charge of 25\,fC was selected for this comparison because it produces a sufficiently large output signal, while it is below saturation for all combinations of the baseline and gain settings. The measurements demonstrate correct pulse shaping and gain programmability at LNT for both
baseline configurations.

\begin{figure}
    \centering
    \includegraphics[width=0.8\linewidth]{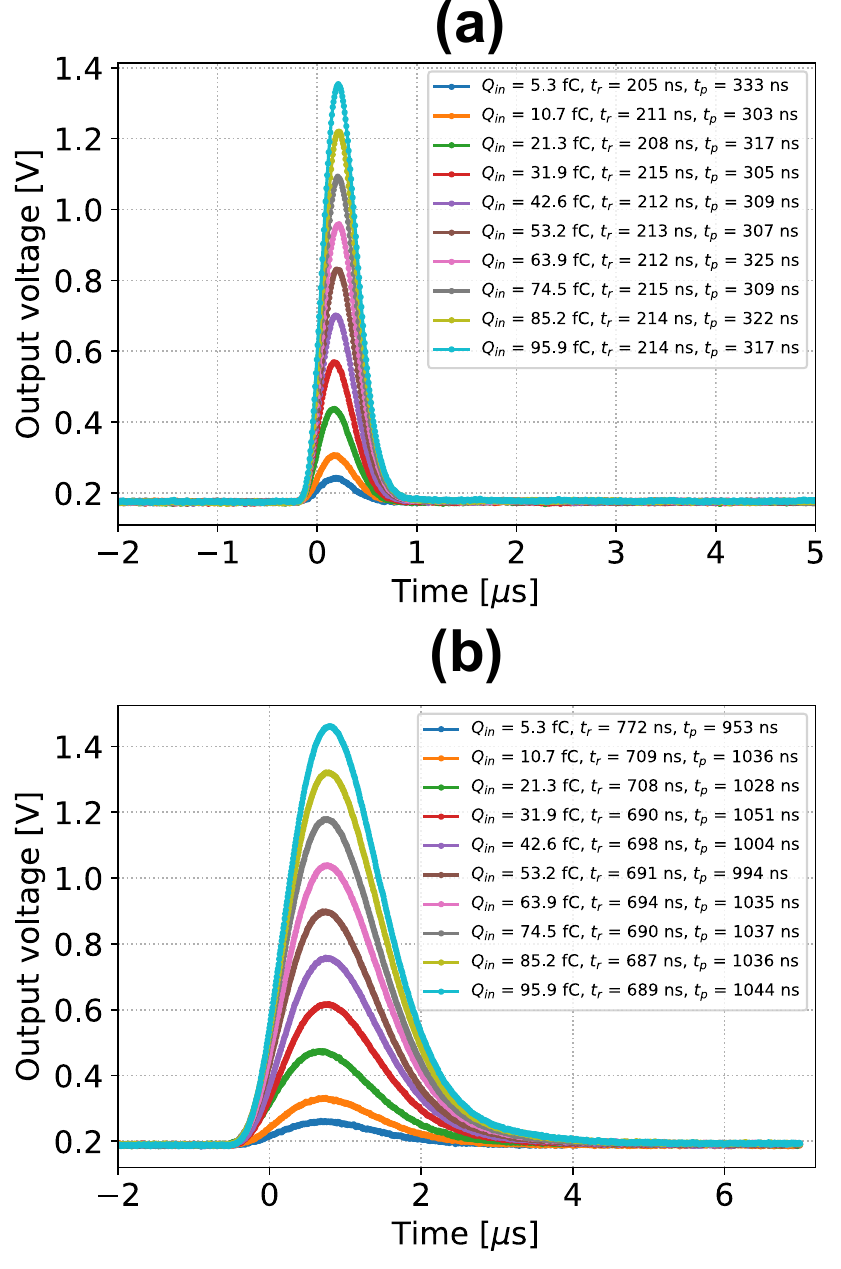}
    \caption{Measured peaking times at LNT for different values of charge injection with baseline = 200\,mV, $g_2$=9, and (a) 250\, ns peaking time setting, (b) 1\,\textmu{}s peaking time setting.}
    \label{fig:LNT-peaking-times}
\end{figure}

The timing characteristics of the shaped pulses were measured at LNT as a function of input charge, as shown in Fig.~\ref{fig:LNT-peaking-times}. The measured peaking time, $t_p$, is defined as the time interval between the points at which the leading edge reaches 5\% of the pulse amplitude and the pulse maximum, while the rise time, $t_r$, is defined as the time interval between 10\% and 90\% of the pulse amplitude. For the 250\,ns peaking time setting, the
measured peaking time ranges from 305 to 333\,ns over the measured input-charge range, with an average value of $\approx$~316\,ns, whereas for the 1\,\textmu{}s peaking time setting, the measured peaking time ranges from 0.95\,\textmu{}s to 1.05\,\textmu{}s, with an average value of $\approx$\,1.04\,\textmu{}s. Both the measured $t_p$ and
$t_r$ values show only a weak dependence on the injected charge.

\begin{figure}
    \centering
    \includegraphics[width=\linewidth]{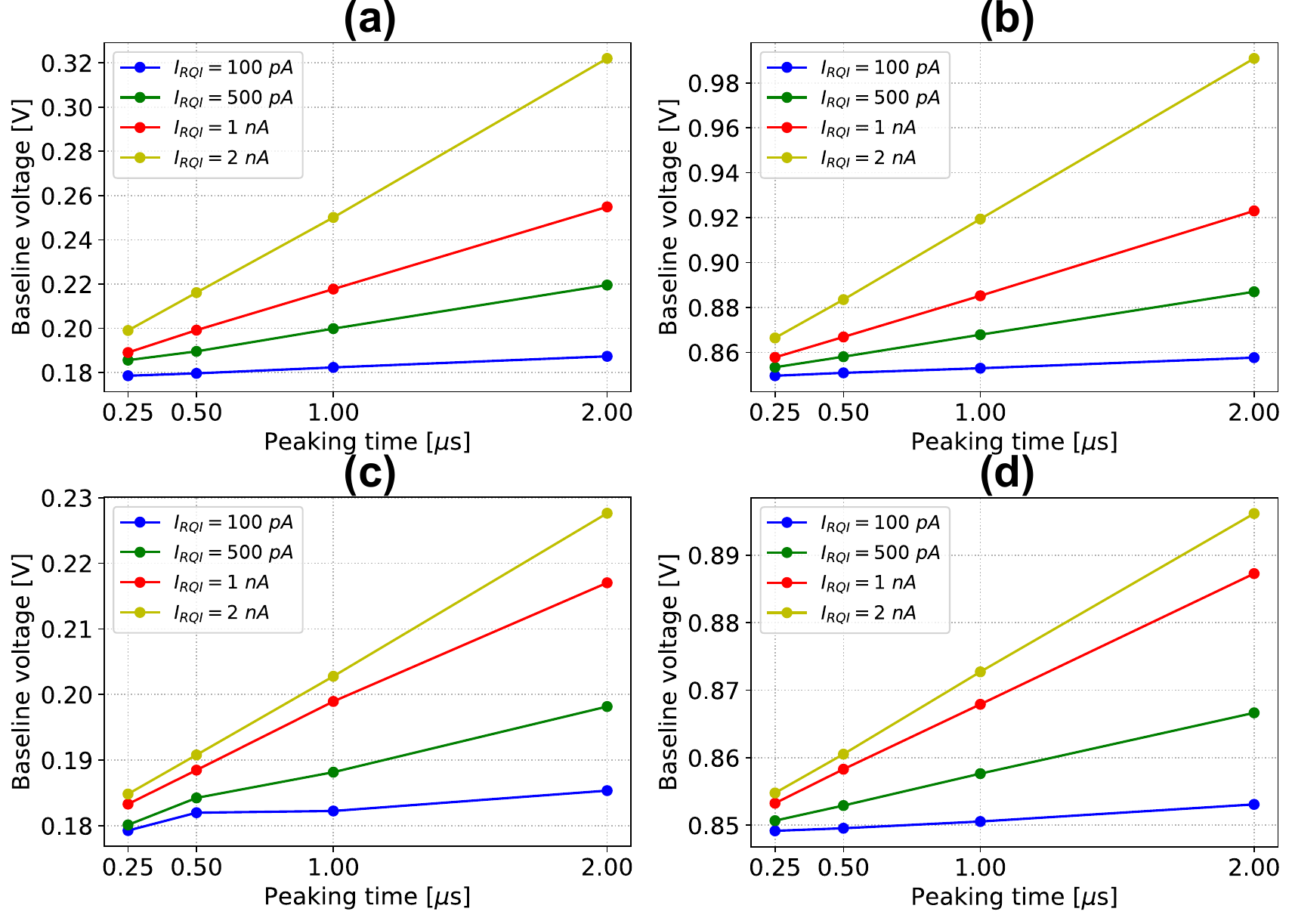}
    \caption{Measured baseline at LNT with gain $g_2 = 9$, and (a) baseline = 200\,mV, no $I_{RQI}$ subtraction, (b) baseline = 900\,mV, no $I_{RQI}$ subtraction, (c) baseline = 200\,mV, $19\times$ $I_{RQI}$ subtraction, and (d) baseline = 900\,mV, $19\times$ $I_{RQI}$ subtraction.}
    \label{fig:LNT-pedestals-with-without-sub}
\end{figure}

\begin{figure}
    \centering
    \includegraphics[width=\linewidth]{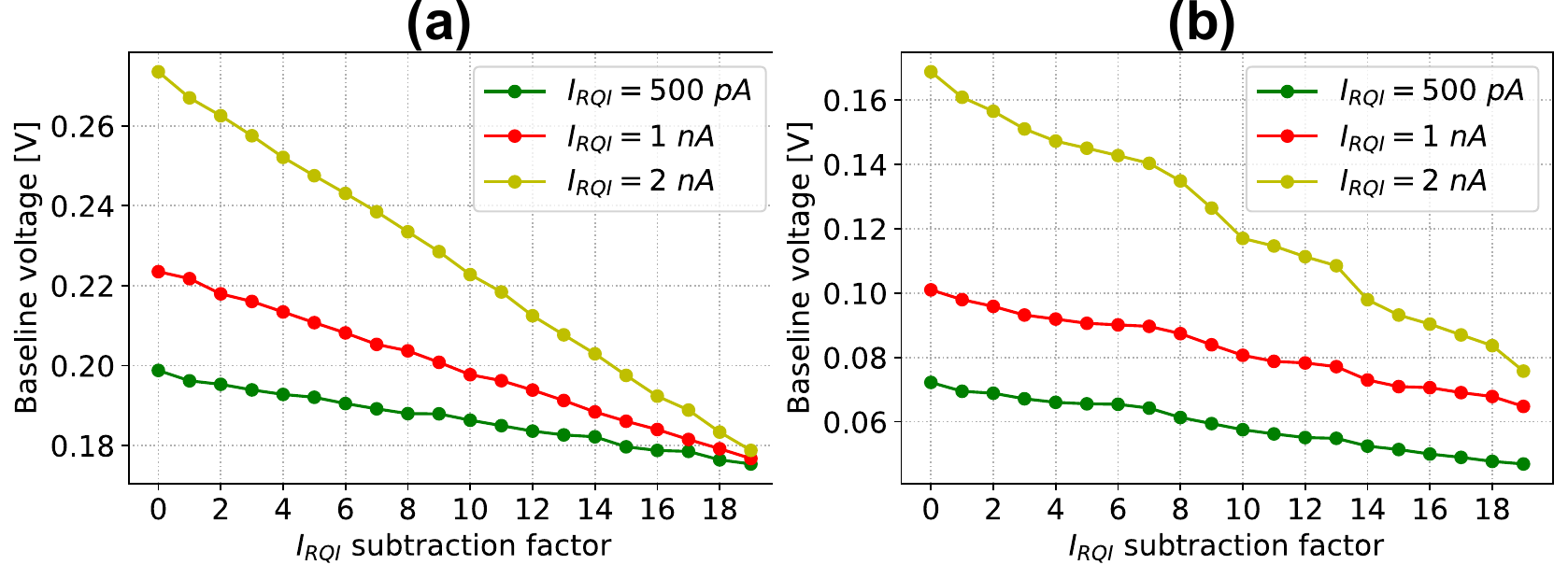}
    \caption{Measured baseline voltage in the 200\,mV baseline mode with gain $g_2 = 9$ and peaking time = 2\,\textmu s at (a) RT, and (b) LNT.}
    \label{fig:baseline-RT-LNT}
\end{figure}

To quantify the effect of $I_{RQI}$ subtraction on the output baseline, measurements were performed at LNT for both baseline settings, with and without $I_{RQI}$ subtraction. As shown in Fig.~\ref{fig:LNT-pedestals-with-without-sub} the maximum baseline change is $\approx$\,140\,mV without subtraction and is reduced to $\approx$\,50\,mV with $19\times I_{RQI}$ subtraction, across the different peaking time and $I_{RQI}$ settings. These measurements demonstrate that the per-channel programmable subtraction circuit effectively restores the available output headroom and also allows equalization of the baseline across channels, thus helping with preservation of the full-scale input charge range,
%restores the available output headroom and the full-scale input charge range, 
particularly for large $I_{RQI}$ values and long peaking times. The monotonic decrease of the baseline as a function of the $I_{RQI}$ subtraction factor is demonstrated by measurements at RT and LNT, as shown in Fig.~\ref{fig:baseline-RT-LNT}(a) and Fig.~\ref{fig:baseline-RT-LNT}(b), respectively. This change in the baseline is set by the current DAC in CA\textsubscript{2} depicted in Fig.~\ref{fig:charms-ca}. It can be observed that the deviations from linearity in the output baseline voltage as the $I_{RQI}$ subtraction factor is varied become more pronounced at LNT. This behavior may be attributed to the increased sensitivity of the weak-inversion M\textsubscript{n3-sub} transistors implementing the current DAC to threshold-voltage mismatch at cryogenic temperatures. In the deep-subthreshold regime, the drain current $I_D$ is approximately given by
$
I_D \propto \exp\left(\frac{q(V_{GS}-V_{TH})}{nkT}\right),
$
where $V_{GS}$ is the gate-to-source voltage, $V_{TH}$ is the threshold voltage, $n$ is the subthreshold slope factor, $k$ is the Boltzmann constant, $T$ is the absolute temperature, and $q$ is the elementary charge. Due to this exponential dependence, the relative variation in the drain current caused by a given threshold-voltage mismatch increases as temperature decreases. Consequently, small threshold-voltage differences among the nominally identical DAC transistors can result in non-uniform subtraction-current increments at LNT, contributing to the observed baseline non-linearity.

% \begin{figure}
%     \centering
%     \includegraphics[width=\linewidth]{figures/linearity_g9.pdf}
%     \caption{Measured linearity at RT and LNT with gain $g_2 = 9$, and baseline = 200\,mV.}
%     \label{fig:lin-g9}
% \end{figure}

% \begin{figure}
%     \centering
%     \includegraphics[width=\linewidth]{figures/linearity_g3.pdf}
%     \caption{Measured linearity at RT and LNT with gain $g_2 = 3$, and baseline = 200\,mV.}
%     \label{fig:lin-g3}
% \end{figure}

\begin{figure}
    \centering
    \includegraphics[width=0.8\linewidth]{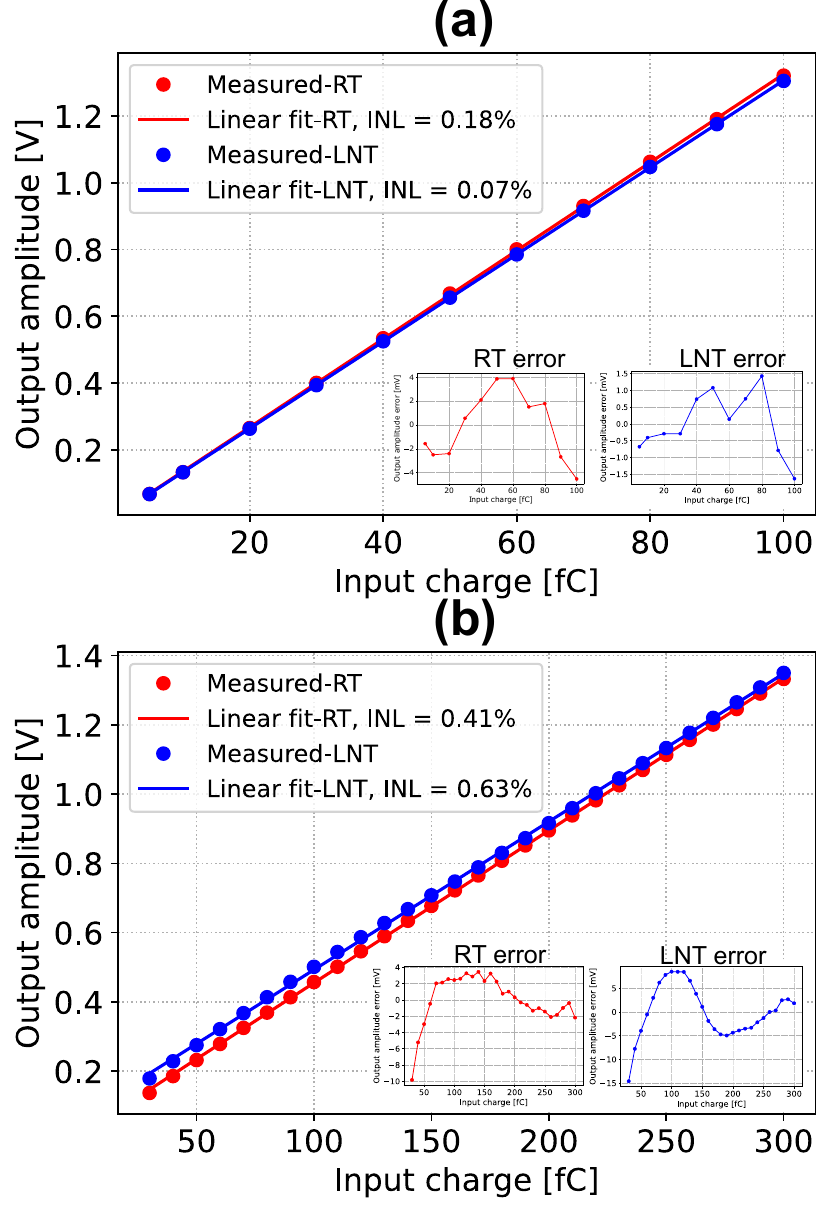}
    \caption{Measured linearity at RT and LNT with baseline set to 200\,mV, peaking time set to 250\,ns, (a) gain $g_2 = 9$, and (b) gain $g_2 = 3$.}
    \label{fig:lin-g3-g9}
\end{figure}

The measured output pulse amplitudes as a function of the injected input charge at RT and LNT for the second stage charge amplifier gain ($g_2$) of 9 and 3 are shown in Fig.~\ref{fig:lin-g3-g9}(a) and (b), respectively. For $g_2=9$, the characteristics of RT and LNT nearly overlap in the measured input charge range, showing
very little difference in gain or linearity at LNT compared to RT. The measured integral non-linearity (INL) is 0.18\% at RT and 0.07\% at LNT. For the lower-gain setting, $g_2=3$, the measured INL is 0.41\% at RT and 0.63\% at LNT, and a larger temperature-dependent change in the charge-to-voltage gain is observed. This behavior is attributed to the signal-dependent resistance and non-linear response of the gain-selection switches in the second charge amplifier stage. The effect becomes more pronounced at LNT due to the increase in the transistor threshold voltage, which reduces the available switch overdrive for large signal excursions. An improved implementation of the gain-selection network in the second charge amplifier stage is planned for the next version of the ASIC to reduce this degradation of linearity, especially at LNT.

\begin{figure}
    \centering
    \includegraphics[width=\linewidth]{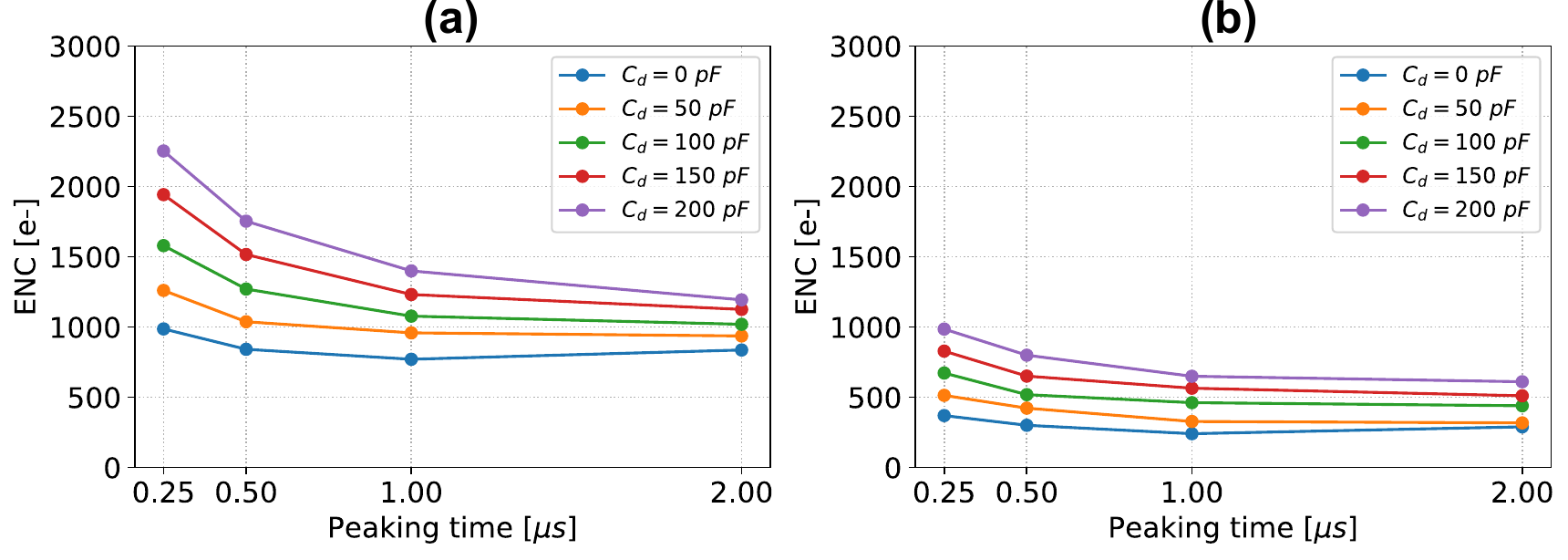}
    \caption{Measured noise in terms of ENC as a function of the peaking time for gain $g_2 = 9$, with baseline = 200\,mV at (a) RT, and (b) LNT.}
    \label{fig:enc_vs_tp}
\end{figure}

\begin{figure}
    \centering
    \includegraphics[width=\linewidth]{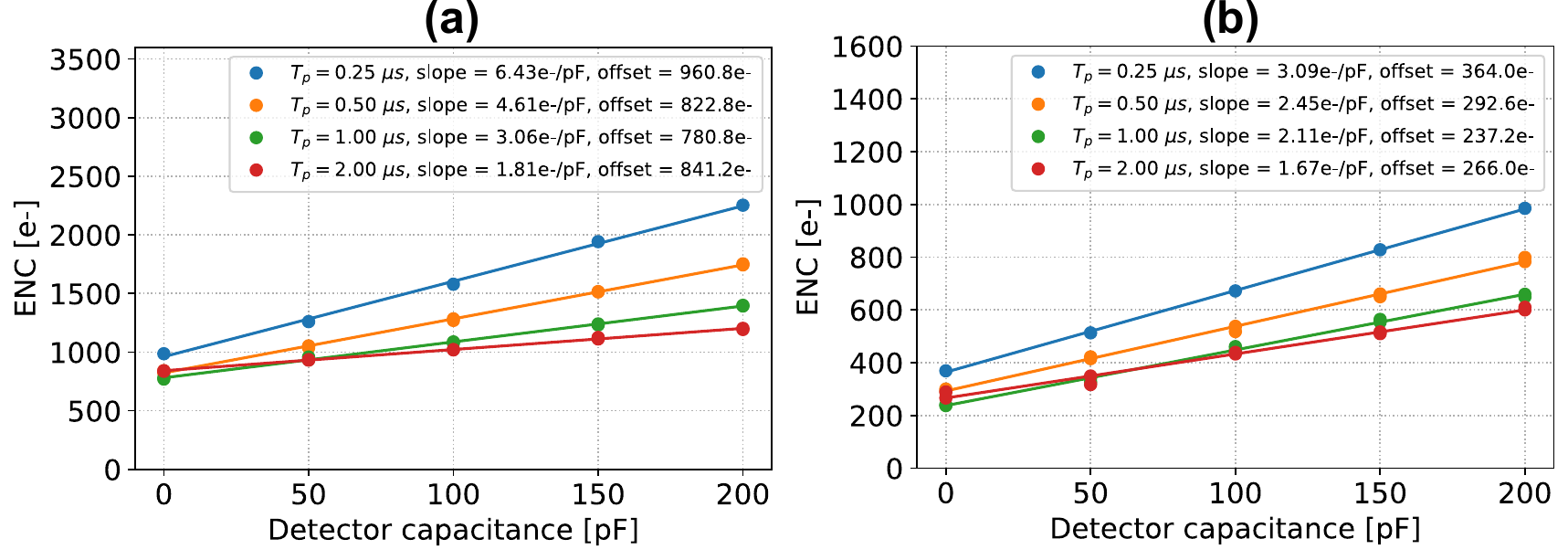}
    \caption{Linear fit of ENC vs. detector capacitance for gain $g_2 = 9$, with baseline = 200\,mV and different peaking times at (a) RT, and (b) LNT.}
    \label{fig:enc_vs_cd}
\end{figure}

The input-referred noise was measured at both RT and LNT for all four peaking-time settings and for detector capacitances, $C_d$, ranging from 0 to 200\,pF. The measured ENC values are shown in Fig.~\ref{fig:enc_vs_tp}. For 200\,pF $C_d$ and $t_p$ of 1\,\textmu{}s, the measured ENC is $\approx$~1400~$e^-$ at RT and $\approx$~650$~e^-$
at LNT, whereas for the shortest peaking time of 250\,ns
the corresponding ENC values are $\approx$~2250$~e^-$ at RT and $\approx~$980$~e^-$ at LNT. Thus, cryogenic operation yields a substantial reduction in ENC. 
 Linear fits to the measured ENC as a function of the detector capacitance for each peaking time setting are shown in Fig.~\ref{fig:enc_vs_cd}. The linear fits provide empirical estimates of the zero-added-capacitance ENC and its increase with detector capacitance over the measured range. The fitted intercept value includes contributions from parallel noise sources as well as the series noise sources associated with the intrinsic input and parasitic capacitances. The slope characterizes sensitivity to detector capacitance and, in the series noise dominated regime, reflects the combined white and 1/f series voltage-noise contributions. At RT, the fitted ENC slopes are 6.4, 4.6, 3.1, and
1.8\,$e^-$/pF for $t_p=0.25$, 0.5, 1.0, and
$2.0$\,\textmu{}s, respectively. Correspondingly, these slopes decrease to 3.1, 2.5, 2.1, and 1.7\,$e^-$/pF at LNT. In particular,
for the shortest peaking time of 250\,ns, each additional picofarad of detector capacitance contributes $\approx$~6.4\,$e^-$ of ENC at RT, while this contribution decreases to approximately half of the RT value, i.e., 3.1\,$e^-$ at LNT. The measured ENC at LNT for CHARMS250V1 is within 10\% of the ENC reported for LArASIC at LNT, for the same value of detector capacitance and peaking time~\cite{TCAS2025}.

\begin{figure}
    \centering
    \includegraphics[width=\linewidth]{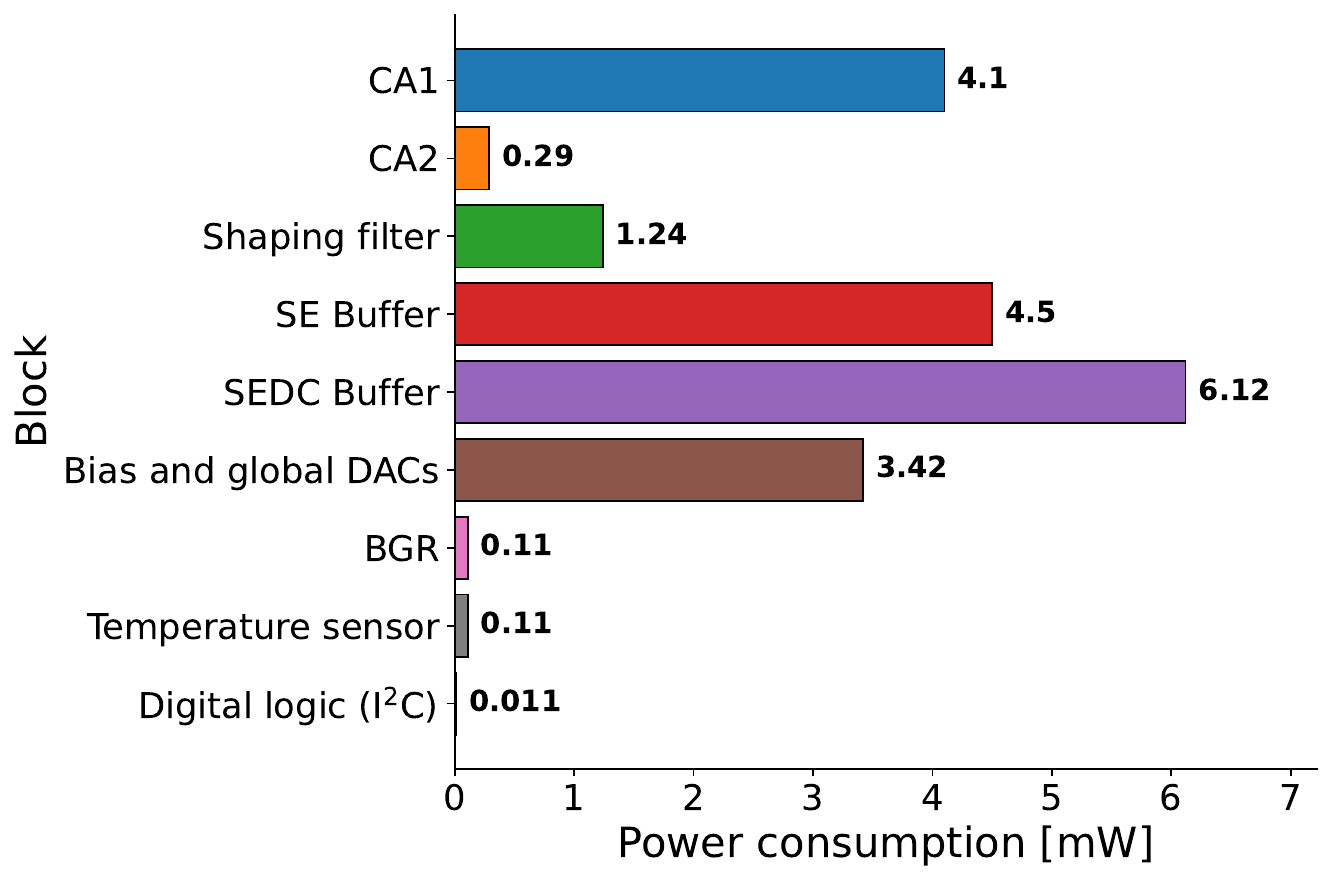}
    \caption{Power consumption breakdown of CHARMS250V1 building blocks.}
    \label{fig:power}
\end{figure}

The measured per-channel power consumption of CHARMS250V1 is 6\,mW when both the SE and SEDC buffers are disabled and the shaping filter output is read out directly, 10\,mW when the output is read out through the SE buffer and 12\,mW when read out through the SEDC buffer. Additionally, auxiliary circuits such as the BGR, temperature sensor, biasing DACs and the I\textsuperscript{2}C block consume a total power of $\approx$~3.6\,mW. The breakdown of power consumption for the channel and auxiliary circuits is summarized in Fig.~\ref{fig:power}.

\section{Conclusion}
\label{sec:conclusion}
This paper presented the design and measurement results of CHARMS250V1, a 65\,nm cryogenic front-end ASIC developed for low-noise readout of charge or light signals in noble liquid detectors. Measurements at 77\,K and 300\,K demonstrate stable operation across the programmable gain, peaking-time, and baseline settings, with less than 1\% integral non-linearity and an ENC of $\approx$~980~$e^-$ at 77\,K for a 200\, pF detector capacitance and 250\,ns peaking time. These results establish the feasibility of implementing low-noise cryogenic front-end electronics in a 65\,nm CMOS technology. Measurements also demonstrate the effectiveness of the programmable reset quiescent current subtraction circuit in reducing baseline shifts and preserving the available output headroom and input charge range. Building on CHARMS250V1, the next generation of CHARMS will integrate 16 front-end channels and is being developed toward the readout requirements of DUNE Phase-II. The architecture will also be extended to provide a substantially larger input dynamic range, targeting the requirements of the noble liquid electromagnetic calorimeter proposed for the FCC-ee ALLEGRO detector concept.

\section*{Acknowledgments}
The authors would like to thank Don Pinelli, Antonio Verderosa, Joe Pinz, and Wenrui (Ray) Lin from the High Density Interconnects Laboratory at Brookhaven National Laboratory for their valuable assistance with wire bonding and test-board assembly, which was instrumental in enabling the experimental characterization and demonstration of the performance of CHARMS250V1.

\bibliographystyle{IEEEtran}
\bibliography{references.bib}

\end{document}